\documentclass[aps,twocolumn,prl,10pt]{revtex4}
\usepackage{epsfig}
\usepackage{amsmath}
\usepackage{amssymb}
\usepackage{graphicx}
\usepackage{latexsym}
\usepackage{rotating}
\usepackage{revsymb}
\usepackage{mathrsfs}

\begin{document}

\renewcommand{\thefootnote}{\fnsymbol{footnote}} 
\renewcommand{\theequation}{\arabic{section}.\arabic{equation}}
\newcommand\degrees[1]{\ensuremath{#1^\circ}}

\title{A genetic algorithm for the atomistic design and global optimisation of substitutionally disordered materials}
\author{Chris E. Mohn}
\affiliation{Department of Chemistry and Centre for Materials Science
 and Nanotechnology, University of Oslo, P.O.Box 1033 Blindern, N-0315 Oslo, Norway}
\author{Walter Kob}
\affiliation{Laboratoire des Collo\"{i}des, Verres et Nanomat\'{e}riaux, UMR 5587, Universit\'{e} Montpellier II-CNRS, 34095 Montpellier, France  }

 \date{\today}
\noindent
  
\begin{abstract}
\noindent

{We present a genetic algorithm for the atomistic design and global optimisation of substitutionally disordered bulk materials 
and surfaces. Premature convergence which hamper conventional genetic algorithms due to problems with synchronisation 
is avoided using a symmetry adapted crossover. 
The algorithm outperforms previously reported Monte Carlo and genetic algorithm simulations for finding low energy minima of 
two simple alloy models {\it without} the need for any redesign.}

\noindent

\end{abstract}
  
\maketitle

\section{Introduction}

Crystal structure prediction and the design of functional materials from first principles 
continue to present a major challenge to the theoretician~\cite{Hawthorne:1990,Mellot:2002,Woodley:1999, Day:2005}. 
It is unfortunate that the presence of imperfections and disorder make these problems even harder.
Nevertheless, taking explicitly into account disorder (both impurities and defect concentrations 
far from the dilute limit), rather than resorting to various mean-field
approaches and simple crystallographic models, is of key importance in order to understand
 material properties at their working temperature and pressure~\cite{Thorpe:2002, Todorov:2004}. 

Many materials with defect concentrations far from the dilute limit
have a substitutional disordered structure. Prime examples involve metallic and semiconductor 
alloys as well as many ceramics, all of which are of tremendous technological importance within numerous fields  as
 electric-, magnetic,- and optical components or devices. 
However, substitutionally disordered materials are
in general not well understood from an atomistic point of view, and describing the 
preferential arrangements 
of the atoms and defects is essential in order to help searching for materials with target 
properties~\cite{Stolen:2006rev}. Identifying the local structural variations within 
disordered compounds is not only essential for 
designing new materials, but the disorder provided by the solid solution of atoms  
in minerals is also essential for understanding the dynamics in the earth's 
interior and of particular importance to assess computationally since these compounds often
only exist at conditions (high pressure and temperature) which are difficult to 
reproduce experimentally~\cite{Lavrentiev:2003}. 

Taking explicitly into account the variation in the local structure is 
not only important for understanding disordered bulk materials, but also essential 
for understanding many surface phenomena. 
For example, simulations of the deposition of small organic molecules 
on various metal surfaces~\cite{Campuzano:1990}, and locating the   
surface-configurations which are likely to influence the 
powder morphology of, e.g., UO$_2$~\cite{Tan:2005} 
are not trivial due to the multitude of different possible atomic arrangements. Also, 
the ability to position atoms at will using a STM-tip~\cite{Tershoff:1985} or 
the opportunities provided by various atomic vapour deposition techniques open the way to efficient 
material design in search for configurations/superstructures with certain target
physical properties  which may be difficult to guess by  chemical intuition.

In a first attempt to attack substitutionally disordered materials by simulation 
in search for, e.g., low energy configurations and superstructures, 
one could carry out an exhaustive and unbiased  
enumeration of all likely candidates. Alas, such a procedure allows only the smallest 
simulation cells to be investigated. Consider, for example, the archetypal 
binary fcc alloy of gold and copper, AuCu. Although the smallest simulation cells can be
investigated by means of brute force enumerations (for instance the small 4 atom
unit-cell has only 12 distinct arrangements), the number of arrangements 
explodes rapidly with the size of the simulation box. 
Within a $2\times 2 \times 2$  simulation cell there are $6 \times 10^8$ ways of arranging the 16 Au and 16 Cu, whereas a 
$4\times 4 \times 4$ (256 atom) supercell contains $6 \times 10^{75}$ configurations. 
Sadly, often much larger cells ($>1000$ atoms) are needed for the  
accurate calculation of many alloy and mineral properties.  
Although symmetry can be used to reduce the search space~\cite{Bakken:2003}, the 
combinatorial explosion of the number of potential energy minima cannot be overcome by 
conventional means, and other strategies are required. 

Finding low energy minima (including the ground state) or configurations with ``key properties'' 
for alloys are in general nondeterministic polynomial-time hard in which heuristics such as 
for instance genetic algorithms (GA) have the potential to be particular suitable~\cite{Holland:1975}. 
Indeed, Smith has demonstrated~\cite{Smith:1992}, in probably the first application of GA for the study of 
disordered materials, that GA are highly efficient in finding low energy configurations for 
several simple alloy-models. It was shown~\cite{Smith:1992} that GA typically outperform the popular Metropolis 
Monte Carlo algorithm (MC)~\cite{Landau:2000}. 

Several contributions have demonstrated how GA can be applied successfully for the study of  
crystalline materials with substitutional disordered structure. Kim~{\it et al}~\cite{Kim:2005} as well as 
Dudiy and Zunger~\cite{Dudiy:2005} applied GA for the study of semiconductor alloys via the use of 
an inverse band structure approach 
in which alloy-configuration with particularly large band-gaps were searched for. 
Applications of GA in search for low energy configurations of simple binary oxide
solid solutions have also been successful~\cite{Mohn:2005gen}. 

However, despite these promising attempts, the dynamics of  genetic algorithms for the 
study of strongly disordered materials is not well understood and problems of choosing a suitable representations 
as well as the role of the different GA-operators require further investigations. In particular 
when bearing in mind that 
recent global energy minimisations of very simple 1D Ising models~\cite{vanHoyweghen:2002} 
(which are very similar to that of finding low energies of alloys) have shown that conventional GA are 
hampered by slow convergence at the late stage of the evolution due to the presence of 
inherent symmetry. We shall therefore take the opportunity to investigate very simple 
alloy models via GA, in an attempt to provide a solution to problems which appear to hamper the use 
of conventional GA for the study of substitutional disorder and related problems in which the 
underyling symmetry may explain the slow convergence of conventional GA.

We follow closely Smith~\cite{Smith:1992}, who carried out detailed tests on several different
simple alloy models and compared the performance of conventional GA with 
that of MC. 
First, we give a general introduction to conventional GA within the field of 
alloy modelling. The basic ingredients are briefly discussed.
Then, having identified and analysed the synchronisation problem which appear to hamper 
conventional GA, we briefly describe an efficient and general GA for finding alloy-configurations 
with target physical properties, e.g. low energy minima, using a simple 1D Ising model as
an example. We discuss the algorithm for two simple 2D alloy models with non-trivial 
representations (genetic encoding), and compare with previously reported GA and MC
simulations. 
The first model is a simple 2D periodic binary ``Ising alloy'' with 
nearest neighbour interactions. This model is an important 
benchmark for testing and analysing the performance of  GA 
since exact analytical solutions exists and since the thermodynamic behaviour 
is well understood.
 The second test is also a 2D alloy model, where the atoms now interact via a Morse potential
with parameters chosen such as to mimic the BiCu alloy. 
This test is particular challenging for conventional GA as 
demonstrated by Smith~\cite{Smith:1992} who found that GA actually perform {\it worse} than 
MC in locating low energy minima. 
We finally compare results from simulations using Ising models and Morse potentials
such as to analyse the influence of the shape of the potential 
(the distribution of fitness) on the performance of GA.

\section{Theory}

\subsection{The alloy models}

\subsubsection{Ising alloy}

The first case is a simple periodic 2D Ising model where the atoms 
are distribution on a square lattice. The potential, originally introduced for the 
study of ferromagnetism in 1925 by Ising~\cite{Ising:1925}, is given as (ignoring external fields)
\begin{equation}
  V = -\frac{1}{2}\sum_{<ij>}J\sigma_i \sigma _j.\label{eq_ising} 
\end{equation}
The ``spins'',  $\sigma_i$, are  either $+1$ (spin up)
or $-1$ (spin down) and $J$ is the (exchange) coupling constant 
between the spins. The summation is carried over
the nearest neighbours only. 
For the study of binary alloys using a potential of the form of equation~\ref{eq_ising}, 
$\sigma_i$ now represent an {\it atom} located  
at site $i$  and $J$ is associated with the bond energy.
We choose $J = -1$, such that bonds between unlike atoms contribute $-1$ to the total energy 
and bonds between like pairs contribute $+1$ to the total energy. The potential is thus designed to 
promote a high degree of mixing between the different species.
For the study of alloys, the summation in equation~\ref{eq_ising} is {\it constrained} to maintain the 
overall composition (or magnetisation). That is
\begin{equation}
m = \sum_{i}\sigma_i = \text{constant},\label{eq_fix}
\end{equation}
throughout the evolution.

\subsubsection{Morse alloy}

Our second test is also a periodic 2D alloy model with atoms distributed
on a square lattice with lattice constant 2.56~\AA. The interactions 
are modelled using a Morse potential of the form
\begin{equation}
V_{ij}(r_{ij}) = D_{ij}[\text{exp}(-2\alpha_{ij}(r_{ij} -\beta_{ij})) -2\text{exp}(-\alpha_{ij}(r_{ij} -\beta_{ij}))].\label{eq_morse}
\end{equation}
Here,  $r_{ij}$ is the interatomic distance between atoms $i$ and $j$. 
$D_{ij}, \alpha_{ij}$ and $\beta_{ij}$ are parameters which are kept fixed for a given interaction 
(see table~\ref{tab_pot}). The parameters were chosen such as
to model the binary BiCu bulk alloy~\cite{Chang:1981} and were employed by Smith~\cite{Smith:1992} 
for benchmarking his genetic algorithm.

\begin{table}
\centering
\caption{Parameters used with the Morse potential.}
\begin{tabular}{cccc}
\hline
\hline
Pot    &  D/eV  & $\alpha$/\AA$^{-1}$ &  $\beta$/\AA~ \\
\hline
 Cu--Cu  & 0.357 & 1.386 & 2.82   \\
 Cu--Bi  & 0.154 & 1.216 & 3.18   \\
 Bi--Bi  & 0.412 & 1.136 & 3.54   \\
\hline\label{tab_pot}
\end{tabular}
\end{table}

In this work we carry out energy calculations using equation 2.3 including 
either only the nearest neighbour interactions, 
or up to the fourth nearest neighbours.

\subsection{Conventional GA for the study of alloys}

Genetic algorithms, originally introduced by Holland in the 1960s~\cite{Holland:1975}, 
are searches for finding good solutions to hard problems.
In contrast to the popular Metropolis Monte Carlo algorithm where changes to 
a single solution are carried out, a population of solutions is evolved. 
Better solutions are found by allowing the ``fitter'' members of the populations 
to {\it recombine} and produce offspring. Random, usually small, changes (mutations) 
can be allowed for, such as to introduce ``fresh blood'' in the population 
and hence avoid problems with premature convergence. The advantage of GA over other 
optimisation algorithms is that large swaps can be carried out in the 
configuration space, 
which is particular useful within the field of global optimisations of ,e.g., 
nano-clusters due to the roughness of the fitness landscape. 
Within this field, there exist reasonable consensus regarding the 
 basic ingredients constituting efficient GA~\cite{Hartke:2004}. However, 
these are 
slightly different compared to those used for the study of periodic alloys 
and related problems such as ferromagnetism. We 
will therefore discuss the GA operations for modelling alloys 
with emphasis on the roles of crossover, mutation, encoding (choice of representation), population sizing,  
fitness etc., and compare in some details the present implementation 
with those of  Smith~\cite{Smith:1992} and Kim {\it et al}~\cite{Kim:2005}.

{\it Encoding:}  
For the purpose of genetic encoding it is useful to distinguish between 
the ``real-space'' solution (phenotype) and  the ``encoded'' solution (genotype).
A convenient choice of representation for modelling binary alloys is a binary string, $\bold x = x_1x_2...x_k...x_N$, where $x_k$ 
is either 0 (if an atom of type A occupy lattice position $k$) or 1 (if an atom of type B occupy lattice position 
$k$), and $N$ is the number of atoms in the simulation box. 
Intuitively, one expect GA to be sensitive on {\it how} the real structure 
is mapped to the form of $\bold x$. Smith~\cite{Smith:1992} analysed the 
sensitivity of the performance of GA in the choice of different  
representations on simple binary alloys, by comparing runs where the atoms were mapped in 
major column form with those of mapping the lattice entirely at {\it random}. 
It was thought that GA may slow down if the positions of the atoms in the string 
does not reflect the actual spatial arrangement of the atoms. 
Although slower, GA were found to perform surprisingly well even if a ``random binary representation'' is used.
Indeed, Kim~\cite{Kim:2005} argue that a non-biased GA should not include any information of the 
spatial arrangements and successfully applied a ``random binary representation''  for the study of bulk alloys. 
By contrast, in ref ~\cite{Mohn:2005gen} a binary representation that strongly 
reflects the spatial arrangements of the crystal structure was successfully 
used for finding low energy configuration of a binary oxide solid solution (the CaO:MgO system). 
This choice of representation seems reasonable 
due to the presence of strong Coulomb forces and the localised nature of the ionic bond. 
In this context, it is worth mentioning that for the optimisation of nano-clusters it is widely accepted
that a real-space representation (which is isomorphic to the phenotype)
performs better than a binary representation, see e.g.~\cite{Johnston:2003}, and 
work is in progress to address these issues in further detail for the optimisation of 
substitutionally disordered materials. 
However, for the optimisation of 2D alloy models we have chosen to use the same 
representation as Smith~\cite{Smith:1992} to allow for direct comparison. 
That is, the 2D lattice is mapped to a 1D binary  
string in major column form, which clearly reflect the spatial arrangements of 
the atoms although more ``compact'' binary representations exists~\cite{Mohn:2005gen}.

{\it Population-size: } 
In GA, the initial population is constructed by selecting configurations at random.
A sufficiently large initial population is in general crucial in order to
avoid premature convergence. However, if the (initial) population is too large, the
efficiency of the GA may slow down. On the other hand, GA are often quite tolerant to changes 
in the population-size. We will discuss the influence of population-size on the 
performance of GA below.

{\it Fitness: }
The members of the population are assigned a ``fitness'' 
which measures the quality of the trial solution. Parents with {\it high} fitness are
preferentially selected for, hereby the popular phrase ``survival of the fittest''. 
The fitness in our work is the total energy (with opposite sign), taken 
either from the simple Ising model (equation~\ref{eq_ising}) or from the 
Morse potential (equation~\ref{eq_morse}). However, as discussed above {\it any} 
physical property can be used.

{\it Selection: } 
There are many different schemes for the selection of parents. 
Popular schemes  are the roulette wheel-, the Boltzmann-  and tournament selections.
In this work we use a binary tournament selection in which two pairs are
randomly chosen, and where the ``fittest'' member in each pair is allowed to mate. 
The binary tournament is convenient, since it does not require any choice of parameter
that controls the selection pressure other than the population-size. 
It is worth mentioning that we have carried out checks with different 
schemes and found that GA are fairly insensitive to the choice of selection method 
but sensitive to the amount of selection~\cite{Mohn:2005gen}.

{\it Crossover: } 
After a pair of parents has been selected, the parents are allowed to mate.
Several types of mating operations (crossover) exist, which are either applied 
on the genotype or directly on the phenotype (the real solution).
For the 2D alloy models, we use a standard binary 
representation where the parents are
encoded in binary form. We use a two-point (TP) crossover, 
$O_{\text{TP}}$: $\mathbb B \otimes \mathbb B \rightarrow \mathbb B \otimes \mathbb B$ ($\mathbb B$ is the configurational space) 
where two different  
integers $k < l \in [1,N] $  are chosen at random, and parents, 
$\bold x = x_1x_2...x_kx_{k+1}...x_lx_{l+1}...x_N$  and $\bold y = 
y_1y_2...y_ky_{k+1}...y_ly_{l+1}...y_N$ are cut between the points $x_k$ and $x_{k+1}$, 
and between $x_l$ and $x_{l+1}$ and spliced as follows
\begin{equation}
  O_{\text{TP}}(\bold x , \bold y) = \bigg \{
  \begin{array}{cc}
    x_1x_2...x_ky_{k+1}...y_{l}x_{l+1}...x_N = \bold v \\
    y_1x_2...y_kx_{k+1}...x_{l}y_{l+1}...y_N = \bold w. \\
  \end{array}\label{eq_cross}
\end{equation}
One of the children, e.g. $\bold v$,  is chosen at random,
possibly mutated (see below), and added to the population by replacing the least fitted member.  
Note that the points where the parents are cut are chosen at
random, but constrained such as to maintain the overall fixed composition. 
This is a severe constraint, and provides a major challenge in designing  
efficient GA for alloy modelling  using single and two-point crossover operators. 
Hence, a {\it uniformed crossover}, where a random set of bits are exchanged between the 
parents, has most frequently been used for the study of periodic alloys 
(see e.g. ref~\cite{Smith:1992}).

{\it Mutation: }
Random, usually small, mutations, $M$,  may be applied typically   
with a small probability ($ < 0.01$).
 In a few cases, we use a permutation operator which exchange a 
pair of different atoms at random positions $k$ and $l$ 
\begin{equation}
 M (\bold z) = M(z_1z_2...z_k...z_l...z_N) = z_1z_2...z_l...z_k...z_N.\label{eq_mut} 
\end{equation}

\subsection{Conventional GA and synchronisation}

Although GA have been highly successful for the study of alloys 
and has the potential to outperform the popular Metropolis Monte Carlo method~\cite{Smith:1992, Kim:2005, Mohn:2005gen}, 
the optimisation of various Ising models via evolutionary algorithms 
have shown that conventional GA are hampered by slow convergence~\cite{vanHoyweghen:2002}.
Simulations carried out in refs~\cite{Smith:1992, vanHoyweghen:2002} have 
shown that the use of highly specialised GA with modified mating schemes and 
rather ``brutal'' mutations (where large blocks of atoms 
were exchanged or scrambled with high probabilities ($>0.1$)) were necessary to find 
global minima. It was argued~\cite{Smith:1992}, that 
such mating operations and mutations are required
due to the constraint of fixing the overall chemical composition which severely limits 
the search-space of the genetic algorithm and leaves very little flexibility for the crossover
to work properly. However, van Hoyweghen {\it et al}~\cite{vanHoyweghen:2002} 
carried out a detailed analysis of the dynamics of GA for finding low energy minima of an 
Ising ferromagnet {\it without} any such constraints and observed a similar slow convergence.

Following van Hoyweghen {\it et al}~\cite{vanHoyweghen:2002}, 
we have analysed the convergence of a genetic algorithm 
for 1D Ising models with periodic boundary conditions. 
Tests were carried out with $J$ = 1 and $J$ =-1 
with and without the constraint of fixing the 
overall composition (or magnetisation if viewed as a spin-problem).
In all these cases, we found, in agreement with ref~\cite{vanHoyweghen:2002}, that 
conventional GA (without mutations) are unable to find global minima of 
a 100 spin Ising model, even with huge population sizes.

\begin{figure}[]
 \begin{center}
 \centering
 \includegraphics[scale=0.45]{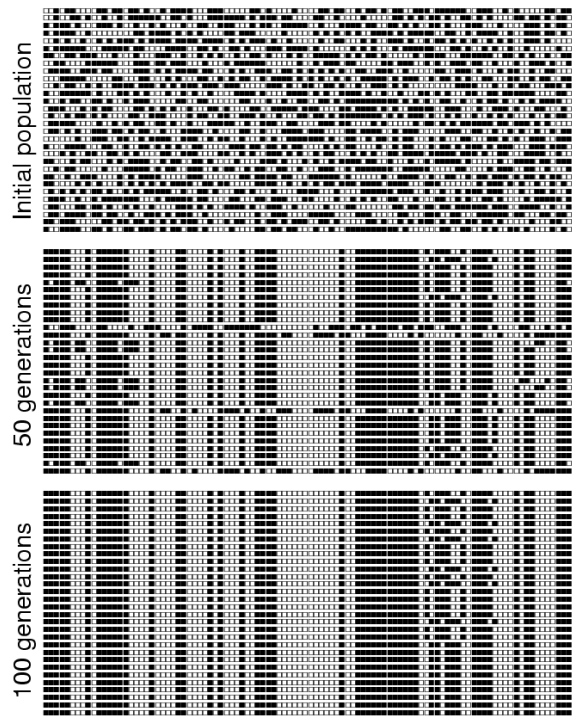}
 \end{center} 
\caption{The populations at three stages using a conventional genetic algorithm applied to a 1D 100
spin Ising model with a fixed 50:50 composition of species A and B. Black and white squares are used to represent
the different species.
Top figure shows the initial population, the middle figure is the population 
after 50 generations (80 fitness evaluations), and the bottom figure shows the population 
after 100 steps (130 fitness evaluations).}\label{fig_conv}
\end{figure}

To visualise the dynamics of a typical conventional genetic algorithm with 
emphasis on the role of the crossover when solving 
Ising-like problems, we show in figure~\ref{fig_conv} snapshots of an evolution 
using a 100 spin Ising model with $J = 1$ and fixed overall composition as an example. 
 A steady state genetic algorithm, where the child replaces the 
worst fit member in the population, is used. 
The population-size is 30 and no mutations were applied.

Although the crossover is very efficient in finding low lying minima
rapidly, the efficiency slows down very early in the evolution. After only 
50 generations the diversity of the population is lost and 
the crossover has become useless. As can be seen, there are no recombinations which 
enables the building blocks within the population to combine to 
form higher order building blocks of an optimium.
 
The failure of conventional GA in finding a global minimum by means of 
recombinations is explained by the loss of a driving force 
due to the high degeneracy of symmetrically equivalent low energy configurations
 which in general are far away from one another in configuration space. 
A modest sized population does not contain sufficient 
diversity to ``differ'' between the building blocks of the low energy minima and the global minima
because the high order building blocks (schemata) of the 
global minima are very similar compared to those of all other low energy minima.
That is, the schemata of the global minima and low energy minima 
contain all high order building  blocks such as **1100**, **1111** and **0000***. 
However, since the number of global minima are outnumbered by the low energy excited states
and since furthermore symmetrically equivalent low energy minima (including the global minima) 
are {\it far away} from one another in the configurational space (in terms of Hamming distance), 
the conventional GA are unable to find global minima, and are stuck in a local optimum.

As can be seen in figure~\ref{fig_conv}, the lack of synchronisation is manifested in the 
representation as domains of **00** and **11** at 
fixed spin-positions  when  conventional GA are used. 
The solutions found are low lying minima indeed, but  
clearly far away from the global minima in Hamming distance. 
Although the performance of conventional GA for attacking different 
Ising models can be improved using specialised operators~\cite{Smith:1992}, 
the strength of the crossover, which should play a major role in designing 
efficient GA, is not fully utilised.

\subsection{Symmetry adapted GA-operators}

Assume that the phenotype possesses some symmetry associated with a group, say 
$\mathscr{G}$, of order $K$, and that the representation chosen is isomorphic to the phenotype. 
Applying a random element $G_a \in \mathscr{G}$ on an arbitrary  
configuration, $\bold x$, leaves the {\it fitness}, $f$, invariant 
\begin{equation}
f ((G_a (\bold x)) = f(\bold x)
\end{equation}
without affecting the ``schemata'' of $\bold x$. At this point, it is worth emphasising, that $\bold x$ now 
represents the real-space configuration rather than the genotype (binary string).  
Replacing, at each generation, the population ($ \bold x, \bold y, ...$) with  
($G_a \bold x, G_b \bold y, ...$), where the symmetry operations $G_a, G_b, ...$
are chosen at random allows all symmetrically equivalent configurations of the
members in the current population to be accessible with equal probability at any step (generation).
Thus, problems due to  synchronisation is avoided, allowing the building-blocks 
of the optima to form also when the evolution is govern by genetic 
drift~\cite{Holland:1975, Goldberg:1989}. 

Turning to our 1D periodic Ising model introduced above, we show how the symmetry can 
be incorporated in a straightforward manner within a conventional genetic algorithm scheme. Note that
the 1D Ising problem, due to periodic boundary conditions, does not have the geometry 
of a string, but of a circle 
\begin{equation}
 \bold x = x_1x_2...x_i...x_N~\text{with}~x_{N+1} = x_1,  
\end{equation}
and hence the encoded representation (genotype) can be trivially chosen isomorphic to the phenotype.
Furthermore, when the overall composition is fixed, the genotype possesses  
the point-symmetry of an $N$-sided polygon and transforms according to the irreducible  
representations of the dihedral group, $D_N$ of order 2$N$. 

Having identified the symmetry of the problem, 
we now introduce a symmetry-adapted crossover (SAC) as follows. 
After  applying the crossover (equation~\ref{eq_cross}), we act with   
a randomly chosen element $G_a \in D_N$ on the child, $\bold x$  
\begin{equation}
 \bold x' =  G_a(\bold x),  
\end{equation}~\label{sym}
and add, instead of $\bold x$, $\bold x'$ to the population.

\begin{figure}[]
 \begin{center}
 \centering
 \includegraphics[scale=0.58]{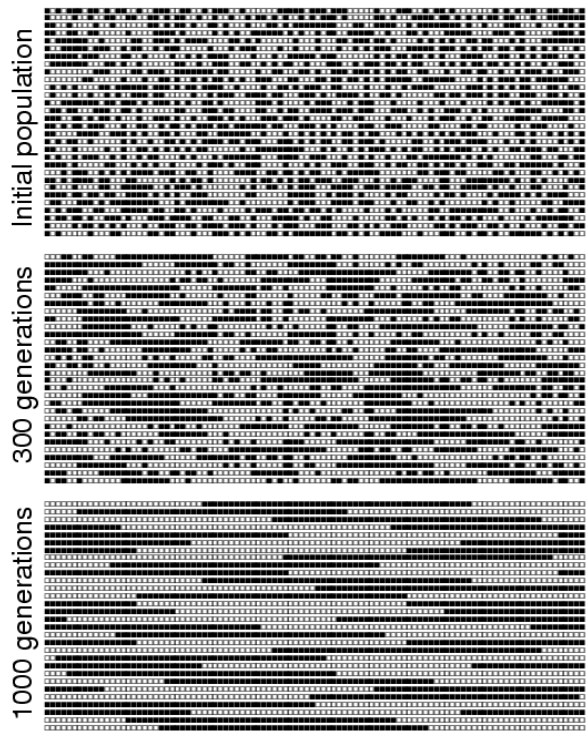}
 \end{center}
\caption{The populations at three stages during the evolution using a genetic algorithm in conjunction with SAC applied to a 1D 100
spin Ising model with a fixed 50:50 composition of species A and B. Black and white squares are used to represent
the different species.
Top figure shows the initial population, the middle figure is the population after 300 
generations (330 fitness evaluations) and the bottom figure shows the population after 1000 steps 
(1030 fitness evaluations).}\label{fig_sym}
\end{figure}

To visualise the dynamics of utilising SAC, we return to our 100 spin Ising model showing 
in figure~\ref{fig_sym} a typical scenario.
As can be seen by comparing the populations after 300 and 1000 steps, 
 the use of symmetry adapted mating operations clearly enables 
the building blocks of good solutions  (e.g. ****1100**** and ****0011****) to combine to 
 form higher order building blocks of even better solutions
 (e.g. **11110000** and **00001111**).
Since the operations $G_a(\bold x)$ leaves the {\it energy} of $\bold x$ invariant {\it without} 
altering the order in which the atoms occur in the representation, large swaps 
can still be carried out in the configurational space when the diversity using  
conventional GA would have been lost (see figure~\ref{fig_conv}). That is, the synchronisation problem is
solved by the use of a phenotype representation and symmetry-adapted operators 
by means of inheriting the symmetry of the problem within the genetic algorithm. Sufficient diversity 
is therefore added, which enables the schemata of good solutions to combine to from schemata 
of even better solutions. 

Now, having presented the use of symmetry-adapted operators, 
we turn our attention to the design of SAC for the study of  
2D alloy models. Since the atoms are distributed 
on a square lattice, the underlying symmetry of the primitive unit-cell 
is that of the plane-group $p4m$. 
Symmetrically equivalent children are thus formed by 
applying the combined operations $G_a T_a $  where $G_a \in  p4m$ and 
$T_a$ is a translation operation. Using a representation which is isomorphic to the 
phenotype involves the use of real-space operations. That
is, cut-and-splice-type operators which are applied directly on the configurations 
rather than via the use of binary encodings. 
However, in order to compare with that of 
Smith~\cite{Smith:1992}, we have chosen to use a representation  
where the atoms were mapped in major column 
form to a binary ring. Since this representation is {\it not} isomorphic to the phenotype, 
the act with a symmetry operations on the children $\bold x'$ may 
alter the order in which the atoms are arranged in the representation. 
However, since this representation strongly reflects the positions of the atoms
in the phenotype  the act of a symmetry-operation will only alter 
very few atoms in the encoded solution and the crossover will work properly.

\section{Results and discussion}

\subsection{Finding low energy minima of a 2D Ising model}

\begin{figure}[]
 \begin{center}
 \centering
 \includegraphics[scale=0.65]{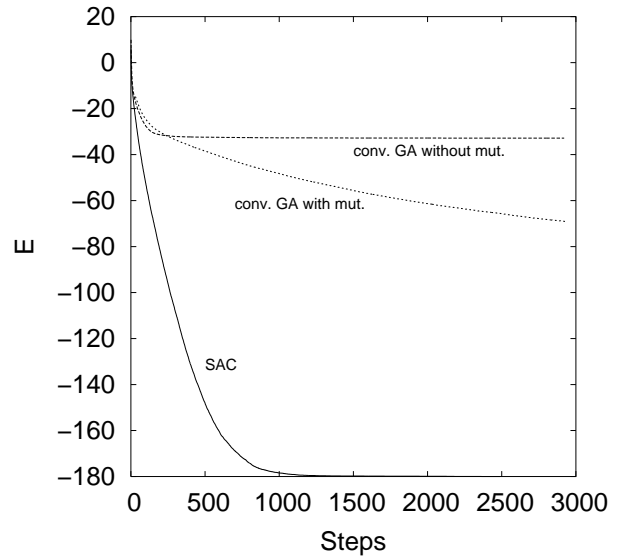}
 \end{center}
\caption{Evolution of the best energy based on 1000 GA runs of a $10 \times 10$ atom Ising model. See the main text for details. The full line is  
that of using SAC, whereas the dotted and dashed lines are GA runs carried out using a conventional genetic algorithm 
with and without mutations respectively. The energy of a global minimum is -180.}\label{fig_ising}
\end{figure}

Results from GA calculations using a 100 atom Ising
alloy with a square unit-cell and equal concentrations of species A and B are shown in figure~\ref{fig_ising}. 
Results from calculations with and without the use of 
symmetry adapted crossover is reported.  In this test $J = -1$
which promotes a high degree of mixing between the different species. 
1000 independent runs were carried out, and the mean value of the best 
energies at each step is displayed in figure~\ref{fig_ising}. 
 A steady state genetic algorithm, where the child replaces the 
worst fit member in the population, is used. 
The population-size is 30. The conventional GA runs were 
carried out both with and without a mutation of exchanging a pair of 
atoms (using equation~\ref{eq_mut}) with probability $0.05$, whereas no
mutations were used during the SAC runs.  

As can be seen from the figure, the use of symmetry-adapted 
operators clearly outperform the conventional GA  
which fails in all attempts to find the global minimum, in agreement 
with previous studies~\cite{vanHoyweghen:2002, Smith:1992}. 
Even with huge population sizes ($>500$) the conventional genetic algorithm is unable to 
find a global minimum without the use of mutation operations. However, as can be seen, 
the convergence when using a simple mutation such as exchanging a
pair of atoms is slow.
By contrast, in all 1000 SAC optimisations, less than 1500 fitness evaluations were required. 
We find that SAC is very robust even when very small populations-sizes (20 members) 
are used, finding global minima in all 1000 GA runs, although the use of larger 
populations ($> 40$ members) slows down the performance. By contrast, if 
the population size is too small ($< 10$ members) GA have the tendencies 
of beeing hampered by premature convergence. 
Using small populations (20 members) is particular advantages when the fitness 
evaluation is expensive 
since a coarse tuning of the parameters can be carried out at relative 
low cost. 

A few test calculations using SAC were also carried out with 
mutations.     
Results from these tests indicate that the use of  mutations 
slows down the performance, although marginally, which is not surprising  
since the diversity of the population is maintained with SAC.

It is worth mentioning that Smith~\cite{Smith:1992} 
is able to find a global minimum using a conventional specialised genetic algorithm with  
a modified mating scheme and large mutations. However, more than 5000 fitness evaluations were typically 
required, which is about an order of magnitude 
slower than that of using SAC, and attempts to solve larger problems
(e.g. a 2D Ising alloy with 400 atoms) failed.   
By contrast, we were able to locate the global minimum 
of a 400 atom Ising alloy in less than 2000 steps. Attempts to find the global minimum 
of Ising alloys with more than 10 000 atoms were successful!

\subsection{Finding low energy minima of a 2D Morse alloy}

\begin{figure}[]
 \begin{center}
 \centering
 \includegraphics[scale=0.65]{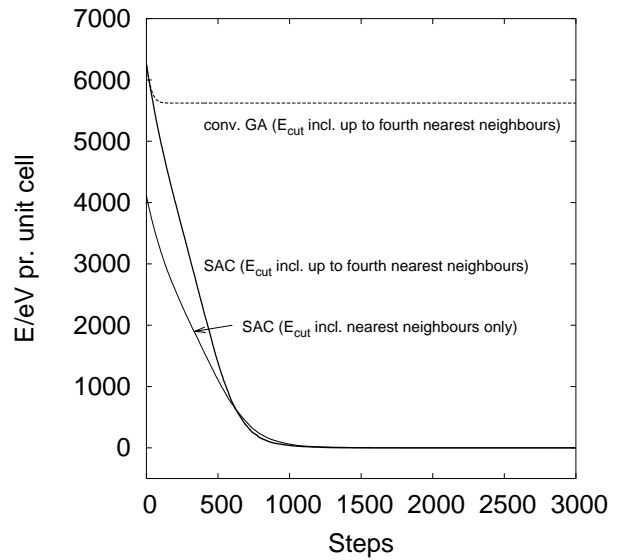}
 \end{center}
\caption{Evolution of the best energy based on 1000 GA simulations of a $10 \times 10$ atom Morse alloy. 
See main text for details. The full lines are  
results from that of using SAC whereas the dashed line is GA runs carried out using a conventional genetic 
algorithm without mutations. The thick full line is the result from runs where up to the forth nearest neighbour interactions were included in the 
energy, whereas the thin full line is result with only  nearest neighbour interactions. The ground state energy is 0 eV in both cases.}\label{fig_morse}
\end{figure}

Having compared the performance of a conventional genetic algorithm with  
that of using SAC for finding low energy minima of a simple Ising model, we now 
turn to discuss the second model where the atoms interact via a
Morse-potential (equation~\ref{eq_morse}) with parameters taken from table~\ref{tab_pot} and
 cutoffs, $E_{\text{cut}}$ in equation~\ref{eq_morse}, chosen such as to the include either nearest neighbour 
interactions or up to the fourth nearest neighbours. Again, a periodic $10\times10$ atom cell is considered 
where 50 Bi and 50 Cu atoms are distributed on a square 
 lattice with lattice constant 2.56~\AA.  
1000 runs were carried out, and the mean value of the best 
energies at each step are displayed in figure~\ref{fig_morse}.  
Results from calculations with and without the use of symmetry adapted crossover is 
reported, and no mutations were used.  

\label{morse}

Again, rapid convergence is achieved using SAC, and less than  1100 steps 
were necessarily in all 1000 runs to 
reach the global minima. The conventional genetic algorithm fails in all attempts,   
hampered by premature convergence.
Smith~\cite{Smith:1992} showed that specialised conventional GA 
when applied for finding low energy minima on a Morse potential 
performed {\it worse} than a Metropolis Monte Carlo algorithm. 
It was argued that the specialised crossover,  which was successful when 
applied to simple Ising
alloys,  fails to work properly when there is a marked difference in the bond energies. 
If the atoms were allowed to 
relax in space the specialised mating operation would work properly because of the similar 
Cu--Cu and Bi--Bi bond energies. Since, in the present work, we do not 
resort to the use of specialised operators, it is of interest to   
compare the performance of SAC with that of Metropolis MC~\cite{Smith:1992}.
Comparing the MC values in figure 7 in ref~\cite{Smith:1992} with that of SAC,  
we find that the MC runs appear to converge faster than SAC early in the evolution. 
However SAC clearly outperforms MC in finding the global minima rapidly.
The high efficiency is due to a properly working crossover throughout the evolution 
which is reflected in 
a steep, roughly linear, decay in the mean energy, followed by an 
exponential decay near the groundstates. 
By contrast, in Metropolis MC changes to a single solution is being made
which is manifested in a different functional form of the mean energy which 
progress at a much slower rate in the latter part of the evolution.

We also carried out test calculations by increasing the cutoff distance in the Morse-potential 
such as to include 
up to the fourth nearest neighbours. Interestingly, as can be seen in figure~\ref{fig_morse}, 
the performance of the genetic algorithm is not very sensitive to 
changes in the cutoff of the Morse-potential which is encouraging for the application of the 
presented genetic algorithm for modelling substitutionally disordered compounds with long-range interactions. 
In addition, the robustness 
of the presented genetic algorithm in handling different alloy models (i.e. with different functional form) 
without having to redesign the algorithm, is particular 
promising for the study of complex minerals and alloys as well as related problems (e.g. magnetism).

\section{Conclusions}

We have presented a novel genetic algorithm for finding configurations with target properties of 
substitutionally disordered materials 
(i.e. alloys and minerals) as well as surfaces in which 
premature convergence hampering conventional GA is avoided.
The presence of a large number of symmetrically equivalent low energy 
configurations, which in general are far away from one-another in configuration space, 
is a challenge to conventional GA since there is {\it no} selection 
in the direction of a single global minimum. The conventional 
genetic algorithm has a synchronisation problem and the crossover fails to work properly when 
the diversity of the population is lost, leaving 
the the problem of finding a global minimum to the mutation operator alone. 

We solve the synchronisation problem using a  
symmetry-adapted crossover in combination with representations which reflect 
the spatial arrangements of the atoms, by replacing the offspring with a
{\it symmetrically equivalent} configuration. The use of symmetry-adapted 
mating operations allows one to carry out large swaps in configuration space, 
enabling the building blocks of good solution to combine to 
form higher order building blocks of even better solutions. 

Calculations on simple 2D alloy models (i.e. an Ising alloy 
and a Morse alloy) show that 
the use of symmetry-adapted operators outperform a conventional genetic algorithm, 
which fails in all attempts to find a global minimum even with huge
population sizes (i.e. with very little amount of selection).
Although Smith~\cite{Smith:1992} is able to find the global minimum 
of a 100 atom Ising model using a specialised genetic algorithm designed to solve 
simple Ising models,  the use of this genetic algorithm fails to outperform Metropolis 
Monte Carlo algorithm when applied to Morse alloys. 
By contrast, we have shown that genetic algorithms in combination with SAC 
outperform Metropolis Monte Carlo-, 
and specialised GA methods in both cases, and global minima of large systems (e.g. alloys with 
more than 10 000 atoms) were found.  

The robustness of the presented algorithm in handling alloy models with different interactions 
{\it without the need for any redesign}, 
is particular promising for future applications.

\section{Acknowledgements}
     
This work was funded by Norges Forskningsr\aa d.  



\end{document}